\documentclass[aps,pre,amsmath,amssymb,twocolumn,showpacs,superscriptaddress]{revtex4}
\usepackage{dcolumn}
\usepackage{graphicx}
\begin{document}
\title{
On the relationship between phase transitions and topological
changes in one dimensional models
}

\author{L.~Angelani}
\affiliation{
        Dipartimento di Fisica,
        Universit\`a di Roma {\em La Sapienza}, P.le A. Moro 2, 00185 Roma, Italy
        }
\affiliation{
        INFM - CRS SMC,  Universit\`a di Roma {\em La Sapienza}, P.le A. Moro 2, 00185 Roma, Italy
        }
\affiliation{
Istituto dei Sistemi Complessi, CNR-INFM, V. dei Taurini 19,
00185 Roma, Italy
}
\author{G.~Ruocco} \affiliation{
        Dipartimento di Fisica,
        Universit\`a di Roma {\em La Sapienza}, P.le A. Moro 2, 00185 Roma, Italy
        }
        \affiliation{
        INFM - CRS SOFT,  Universit\`a di Roma {\em La Sapienza}, P.le A. Moro 2, 00185 Roma, Italy
        }
\author{F.~Zamponi} \affiliation{
        Dipartimento di Fisica,
        Universit\`a di Roma {\em La Sapienza}, P.le A. Moro 2, 00185 Roma, Italy
        }
  \affiliation{
        INFM - CRS SOFT,  Universit\`a di Roma {\em La Sapienza}, P.le A. Moro 2, 00185 Roma, Italy
        }

\date{\today}
\begin{abstract}
We address the question of the quantitative relationship between
thermodynamic phase transitions and topological changes in the
potential energy manifold analyzing two classes of one
dimensional models, the Burkhardt solid-on-solid model and the 
Peyrard-Bishop model for DNA thermal denaturation, 
both in the confining and non-confining version.
These models, apparently, do not fit [M. Kastner, Phys.
Rev. Lett. {\bf 93}, 150601 (2004)] in the general idea that the
phase transition is signaled by a topological discontinuity. 
We show that in both models the phase
transition energy $v_c$ is actually non-coincident with, and
always higher than, the energy $v_\theta$ at which a topological
change appears. However, applying a procedure already successfully
employed in other cases as the mean field $\phi^4$ model, i.~e.
introducing a map ${\cal{M}}\!: v\to v_s$ from levels of the
energy hypersurface $V$ to the level of the stationary points
"visited" at temperature $T$, we find that ${\cal{M}}
(v_c)$=$v_{\theta}$. This result enhances the relevance of the 
underlying stationary points in determining the
thermodynamics of a system, and extends the validity of
the topological approach to the study of phase transition to
the elusive one-dimensional systems considered here.
\end{abstract}

\maketitle

Phase transitions are a very well understood subject
in statistical mechanics. They have been characterized in many different
ways in the last century and many observables related to the phase
transition ({\it e.g.} critical exponents, correlation lengths, etc.) have
been computed and measured with very high accuracy \cite{Ma}.

Recently, a new characterization of phase transitions has been proposed
by Pettini and coworkers \cite{cccp,fps,cpc,fra_pet}.
These authors conjectured that, for classical systems defined by a continuous 
potential energy function $V(\{q_i\}_{i=1\ldots N})$,
a thermodynamic phase transition, occurring at a temperature $T_c$, 
is the manifestation of a topological discontinuity, taking place at a
specific value $v(T_c)=N^{-1} \langle V(q) \rangle$ (where $\langle \cdot \rangle$ is the
statistical average at temperature $T$) of the potential energy function $V$,
or, more precisely, taking place on the hypersurface $\Sigma_v =\{
(q_1,\dots,q_N)\in{\Bbb R}^N \vert\ V(q_1, \dots ,q_N) = Nv\}$, at $v=v(T_c)$.
The most striking consequence of this hypothesis is that the signature of 
a phase transition is present in the topology of the configuration
space independently on the statistical measure defined on it.

The changes in the topology are identified through the Morse
theory \cite{morse}: according to this theory the topological
changes in a manifold like $\Sigma_v$ are related to the
presence of stationary points of $V$ (points for which $\nabla
V$=$0$) at energy $v$. 
However, the precise meaning of the correlation between topological
changes and phase transitions in the general case is still a open question. 
From one hand there
is a theorem of Franzosi and Pettini \cite{fra_pet}, asserting
that, for ``smooth, finite-range and confining microscopic
interaction potentials $V$ with continuously varying
coordinates,\dots, a topology change of the 
$\{\Sigma_v\}_{v\in{\Bbb R}}$ at some $v_{\theta}$ is a {\it
necessary} condition for a phase transition to take place at the
corresponding energy \dots value'' \cite{fra_pet}. On the other
hand, there are different numerical studies of various models
\cite{fps,XY,ktrig,ktrig2,phi4,kastner,gri_mos,kastner2,teix}
(almost all with potentials $V$ {\it not} fulfilling the
hypotheses of the theorem) for which a variety of results has been
obtained: some are in agreement with the ``{\it topological
hypothesis}'' of Pettini and coworkers, others seem to indicate
its failure. 

It is important to underline that the theorem in 
Ref.~\cite{fra_pet} establishes a {\em necessary} condition for a phase
transition to take place. The problem to find {\em sufficiency}
conditions is still an open problem, as pointed out also by the
proponents of the hypothesis. 
This question has been addressed in two recent papers: in Ref.~\cite{kastner}  
a one-dimensional model, the Burkhardt model with nonconfining
potential (see below) was investigated, finding that
a topological change is present without a phase transition at finite temperature;
in Ref.~\cite{teix} the mean field spherical ferromagnet was considered, and it was found
that the same topological changes happen either in absence and in presence of a magnetic
field, while in the latter case no phase transition occurs
(it is worth noting, however, that these two models do not fulfill the
hypotheses of the theorem in Ref.~\cite{fra_pet}).
Moreover, in the thermodynamic limit, it is likely that, for any finite
interval of energy $I=[v_1,v_2]$, there is always a stationary point
of $V(q)$ with energy $v \in I$; thus, in the thermodynamic limit
a topological change occurs with probability one in any finite interval 
of energy.
Most of these topological changes are --obviously--
not related to phase transitions,
and indeed it seems that a topological change must be {\em strong enough}
to be related to a thermodynamic phase transition
(see Ref.~\cite{XY} for a detailed discussion of this point).
An indication coming from the analysis of the models cited above is
that the presence of a phase transition should be related to the 
presence of a {\em singularity} in the Euler characteristic at 
a given energy $v_\theta$.
Thus, basically, the idea of the ``{\em topological}''
approach is that the phase transitions are correlated to abrupt 
changes in topological
quantities defined on the stationary points of $V$ (as, for example,
the Euler characteristic).

Another open question is the equivalence between the energies
at which phase transition ($v_c$) and change in the topology
($v_{\theta}$) take place. The original conjecture of Pettini
and coworkers asserts that the two energies ``{\em correspond}'' 
(let's call this the {\it strong topological hypothesis}). 
To our knowledge there is only one system within the hypotheses 
of the theorem, the two dimensional $\varphi ^4$ model \cite{fps}, 
for which
the equivalence has been numerically established. In other two
systems with long range interactions, thus out of the theorem
hypotheses (the mean field $XY$ model \cite{XY} and the 
mean-field $k$-trigonometric model \cite{ktrig,ktrig2}) the equivalence has been
analytically proved. There are, conversely, analytical results for
a different model system (mean field $\varphi ^4$ model
\cite{phi4}) for which the correspondence does not hold: the
energy $v_c$ at which phase transition takes place is higher than
the energy $v_{\theta}$ of the topological singularity
(we stress here that also in this case the hypotheses of the theorem
in Ref. \cite{fra_pet} are not fulfilled).

The latter discussion is of particular importance, not only in the
context of phase transitions, but also in analogy with
glassy systems. These systems are characterized --at the mean field level--
by a dynamical transition taking place at a given temperature $T_{_{MCT}}$ (or
equivalently at energy $v_{_{MCT}}$) predicted by mode-coupling
theory, and a (hypothesized) true phase transition at a lower
temperature $T_K$ (Kauzmann temperature) or energy $v_K$. From
numerical simulations of Lennard-Jones like systems, one observes
that the dynamical transition is strictly related to the
properties of the saddles visited by the system \cite{sad_lj,sad_cav}. 
The concept of ``visited saddles'' is quantitatively worked out defining a
pseudo-potential $W$=$|\nabla V|^2$ and minimizing it during the
dynamic evolution of the system, thus obtaining a map
${\cal{M}}_q:q \rightarrow q_s$ associating to each equilibrium
configuration point $q=\{q_1,\dots,q_N\}$ a minimum $q_s$ of $W$.
When averaged over the dynamic trajectory one obtains an energy
map: ${\cal{M}}:v \rightarrow v_s$. Absolute minima of $W$ (having
$W(q_s)$=$0$) correspond to stationary points (saddles and minima)
of $V$. We note that the presence of local minima of $W$, with
$W(q_s)\neq 0$ but small (corresponding to inflection directions
in $V$ profile), does not affect the result \cite{jcp_comment}: 
the order of visited
saddles (number of negative eigenvalues of the Hessian matrix of
$V$) extrapolates to zero at $T_{_{MCT}}$, and the energy of
saddles stays always below the instantaneous energy. Moreover the
true thermodynamic transition is achieved when the number of
visited stationary points of order zero (minima of $V$) grows less
than exponentially with the system size (in the glassy terminology
when the ``{\em configurational entropy}'' or ``{\em complexity}''
goes to zero). Solvable mean field spin-glass models ($p$-spins),
which manifest the same phenomenology of structural glasses,
corroborate these findings in an analytical way \cite{cavagna}.
Then, what emerges from glassy systems, is the great importance of
the {\em underlying} stationary points in the description of the
various transitions (dynamical and thermodynamical) taking place
in these systems. It is worth to note, however, 
that the consistency of this picture beyond
mean field is still matter of debate \cite{VMF}, and that the definition
of the map ${\cal M}$ is not unique also at the mean field level,
different definitions giving similar but not quantitatively equal
results \cite{phi4}.

One can argue, in line with the ``{\em
topological}'' approach to phase transitions, that also for
non-glassy systems the concept of {\em underlying} stationary
points continues to be useful. It is important to emphasize that,
in the study of the glass transition, is the discontinuity of the
average density number of underlying stationary points that marks the
dynamical transition at $T_{_{MCT}}$. Driven by this observation,
we recently proposed that the map ${\cal{M}}:v \rightarrow v_s$
has to be applied in order to spot the phase transition, i.e., if
a topological discontinuity exists at energy $v_{\theta}$, the
phase transition is expected at an energy $v_c$ such that ${\cal
M} (v_c)$=$v_{\theta}$. This has been proved to work 
(at least approximately) in
those cases (e.g. the mean field $\varphi ^4$ model) where the
original ``{\em strong topological hypothesis}'' (i.e. coincidence
between $v_c$ and $v_{\theta}$) failed. It is worth to point out
that those cases where it has been proved that $v_c \equiv
v_\theta$ do not constitute counterexamples for the application of
the map ${\cal M}$, as in all these cases it turns out that $v_c$
is a fixed point for the map, i.~e. ${\cal M} (v_c)$=$v_{c}$. In
conclusion, for all the cases investigated so far \cite{XY,ktrig,ktrig2,phi4}, it
results that whenever a phase transition (including also "dynamic"
transitions as the glass transition in LJ liquids \cite{sad_lj,sad_cav} 
and p-spin systems \cite{cavagna}) 
is present at a certain energy $v_c$,
this transition is signaled by a discontinuity in the topology,
specifically in the Euler characteristic or in the complexity, at
an energy $v_\theta$ such that ${\cal M} (v_c)$=$v_{\theta}$. At
variance with the original ({\it strong}) topological hypothesis,
where it was supposed that $v_c=v_\theta$, we will refer to the
latter conjecture as {\it weak topological hypothesis}.
Note that the {\it weak topological hypothesis}, 
at variance with the {\it strong topological hypothesis},
depends on the statistical measure, as the map 
${\cal M} : v \rightarrow v_s$ is defined through an {\it average}
over the dynamical trajectory (or, equivalently, over the statistical
measure). We will discuss this point in detail in the following.

Two recent papers \cite{kastner,gri_mos} addressed the question
concerning the relationship between phase transitions and topology
in one dimensional models. Kastner \cite{kastner} studied
two versions of a solid-on-solid model, one showing a phase
transition at finite temperature and the other not; he found that
both models exhibit the same topological change, thus concluded
towards an ``unattainability of a purely topological criterion for
the existence of a phase transition'' \cite{kastner}. Grinza and
Mossa \cite{gri_mos} considered the Peyrard-Bishop model
\cite{pey,pey1}, which exhibits both a phase transition and a change in
the topology, but in this case $v_c$ and $v_{\theta}$ are not
coincident \cite{nota}. These papers contributed to extend the
analyzed cases for the understanding of necessary and sufficient
conditions for the {\em topological hypothesis}. However, they
seem to reach contradictory results, one supporting and the other
falsifying the topological hypothesis, even if the investigated models
share many similarities.

The aim of this work is to try to clarify this apparent inconsistency
with regard to what has been discussed above.
In particular, we reanalyze the
model investigated by Kastner and by Grinza and Mossa. As a result
of this study {\em i)} analyzing the Peyrard-Bishop Model (PB)
\cite{pey,pey1} we numerically show that the two energies, although
different $v_c\neq v_{\theta}$, satisfy the ``{\em weak
topological hypothesis}'', i.e. ${\cal M}(v_c)$=$v_{\theta}$. We also
investigate a slight modification of the PB model (allowing
non-confined motion of the variables), where we are able to study
the same quantities in absence of a thermodynamic phase transition
at finite temperature,
again finding results in agreement with the ``{\em
weak topological hypothesis}''. 
{\em ii)} Analyzing the Burkhardt
model we introduce a further parameter defining the position of
the pinning potential, that can be moved from the origin, i.e.
fully confining potential, to infinity, fully non-confining
potential. 
We found that the phase transition actually exists for
{\it all} the position of the pinning potential, and its critical
temperature goes continuously to infinity as the pinning potential
position goes to infinity.
Moreover, we found
that also the generalized Burkhardt model falls into the class of
systems that satisfy the ``{\em weak topological hypothesis}'',
i.e. ${\cal M}(v_c)$=$v_{\theta}$.
As the position of the pinning potential is moved toward
infinity, the energy $v_s(T)$ of the ``{\it underlying saddles}'' 
tends to reach the energy $v_\theta$ at higher temperature;
when the nonconfining limit is reached, $v_s(T) \leq v_\theta$ for
all $T$. The topological singularity is visited only for $T \rightarrow \infty$
and this is the reason why the phase transition is not observed,
i.e. $T_c \rightarrow \infty$.

The present findings support the idea that also in the case of one
dimensional models, the relevant topological
quantity related to a phase transition is obtained from underlying
stationary points obtained from the map ${\cal M}$. As we already
noted, the choice of the map ${\cal M}$ is not univocal: we
have chosen the one obtained through $W$ (with some {\em ad hoc}
modifications), however different
choices are possible (we mention here, for example, the map
obtained using Euclidean distances in configuration space
\cite{phi4,cavagna}). The robustness of this conclusion with respect to
the possible different choices of the map ${\cal M}$ is still an
open question which goes beyond the scope of this paper.

\section{The models}
\begin{figure}[t]
\includegraphics[width=.5\textwidth]{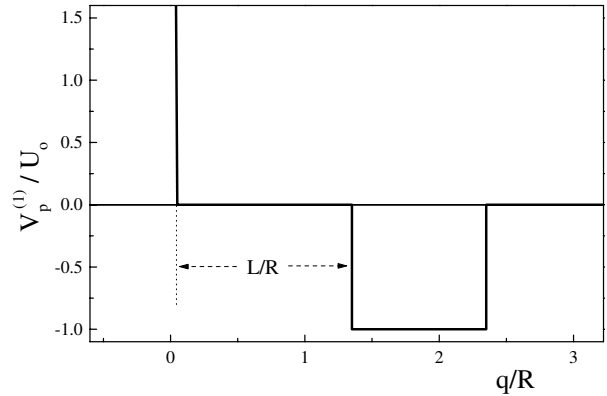}
\caption{Sketch of the on-site pinning potential $V^{(1)}_{p}(q) $
for a given choice of the control parameter $L/R$. } \label{fig_1}
\end{figure}

The one dimensional models we study are all defined by the
Hamiltonian ${\cal H} = \sum_{i=1}^{N} p_i^2/2m +
V(\{q\}_{i=1\dots N})$ ($m$ is the mass of each particle), where $V$
is the potential energy. We consider two different classes of
models. The first one, introduced by Burkhardt \cite{bur} as a
model for localization-delocalization transition of interfaces, is
defined by the potential energy $V^{(1)}$:
\begin{equation}
V^{(1)}(\{q\}_{i=1\dots N}) = \sum_{i=1}^{N} K |q_{i+1}-q_i| +
\sum_{i=1}^{N} V^{(1)}_{p}(q_i) \ , \label{v_bur}
\end{equation}
where $K$ measures the strength of the force between neighboring pairs,
$V^{(1)}_{p}(q)$ is the on-site pinning potential,
and periodic boundary conditions are assumed $q_{N+1} \equiv q_1$.
We chose for $V^{(1)}_{p}(q)$ the following form
\begin{equation}
V^{(1)}_{p}(q) = \left\{
\begin{array}{cl}
+ \infty \ \ \ \ \ & \mbox{for $q \leq 0$}
\\
0        \ \ \ \ \ & \mbox{for $0 < q < L$}
\\
 -U_0    \ \ \ \ \ & \mbox{for $L \leq q \leq L + R$}
\\
0        \ \ \ \ \ &  \mbox{for $q >  L + R$}
\end{array} \right.\label{gburk}
\end{equation}
that generalizes the original form in Ref. \cite{bur} introducing
a parameter $L$ that gives the position of the pinning potential
(a square well of depth $U_0$ and width $R$, see Figure \ref{fig_1})
from the edge of the system. The case with $L=0$ coincides with
the original Burkhardt confining model, while the non-confining
case is retrieved in the $L\rightarrow \infty$ limit.

\begin{figure}[t]
\includegraphics[width=.5\textwidth]{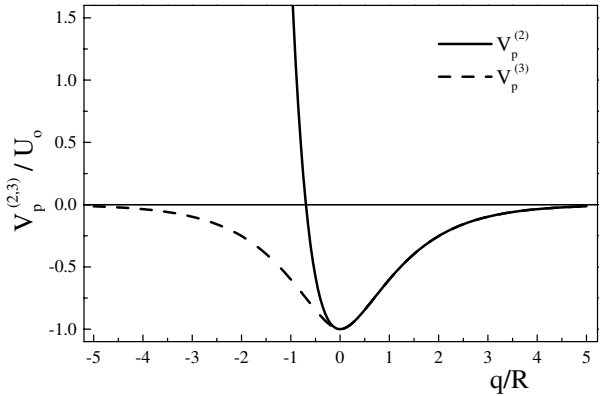}
\caption{Plots of the on-site pinning potentials $V^{(2)}_{p}(q)$
(full line) and $V^{(3)}_{p}(q)$ (dashed line). } \label{fig_2}
\end{figure}

The models of the second class are defined by the the potential energy
$V^{(2)}$ and $V^{(3)}$ of the form:
\begin{equation}
V^{(2,3)}(\{q\}_{i=1\dots N}) = \sum_{i=1}^{N} \frac{K}{2}
(q_{i+1}-q_i)^2 + \sum_{i=1}^{N} V^{(2,3)}_{p}(q_i) \ .
\label{v_pb}
\end{equation}
We consider two different versions of this model, one defined by
the on-site Morse potential, introduced by Peyrard and Bishop
as a simple model for DNA thermal denaturation
\cite{pey,pey1} (PB model)
\begin{equation}
V^{(2)}_{p}(q)  = U_o \{ (e^{-q/R} -1)^2 -1\}\ ; \label{model1}
\end{equation}
the other is a symmetric version of the former (SPB model)
\begin{equation}
V^{(3)}_{p}(q)  = U_o \{ (e^{-|q|/R} -1)^2-1\} \
, \label{model2}
\end{equation}
a slight modification of PB model that allows for a non-confined motion
of the variables (see Fig. \ref{fig_2}). 
We note that the introduction of the modulus in
the Eq. \ref{model2} does not introduce discontinuities up to the
second derivative of the potential. The quantities $U_o$ and $R$
determine respectively the energy and the length scales of the
on-site potential (in the following all quantities will be
reported in $U_o$ and $R$ units). We further chose $m$=1. In all
the three cases a parameter of the Hamiltonian is related to the
strength of the inter-particles interactions ($K$) and, for the
case of $V^{(1)}$ a second parameter is the position of the
pinning potential $L$.

Specifically, the relevant quantity defining the relative weight
of the on-site with respect to interparticles potentials is the
dimensionless ratio $\xi$=$K R/U_o$ or $\xi$=$K R^2/2U_o$ for the
potential models (1) or (2,3) respectively, while the position of
the pinning potential is given by $\zeta=L/R$ 
for the potential model (1).

The generalized Burkhardt model has been treated analytically,
while the Peyrard-Bishop models are studied numerically. In the
latter cases we performed isothermal molecular dynamics
simulations using Nos\'e-Hoover thermostat at different
temperatures for systems with $N$=$500$ degrees of freedom with
periodic conditions $q_{N+1}$=$q_1$. We studied different values
of the control parameter $\xi$, as an example the results are
reported for $\xi$=$0.05$ and $0.5$, all the other $\xi$ values give
results in qualitative agreement with the two reported examples.

\section{Burkhardt model}

\subsection{Thermodynamics}
The thermodynamics of the Burkhardt model is known since many years
\cite{bur} for both the $\zeta=0$ and $\zeta=\infty$ cases. The
method of solution is briefly outlined below.

The determination of the thermodynamic of systems described by a
potential function of the form
\begin{equation}
V(\{q\}_{i=1\dots N}) = \sum_{i=1}^{N} K |q_{i+1}-q_i| +
\sum_{i=1}^{N} V_{p}(q_i) \ , \label{forma}
\end{equation}
i.~e. similar to the case (1) (Eq.~\ref{v_bur}), goes through the
exploitation of the transfer matrix technique. Indeed, the
configurational partition  function ${\cal Z}$ is given by:
\begin{equation}
{\cal Z}_N = \int dq_1\dots dq_N \; e^{-\beta V(\{q\}_{i=1\dots N})} \ ,
\label{Z}
\end{equation}
that, defining the transfer "matrix"
\begin{equation}
{\cal T}(x,y) = e^{-\beta K \vert x-y\vert} \; \; e^{-\beta
[V_{p}(x)+V_{p}(y)]/2}  , \label{T}
\end{equation}
can be written as:
\begin{equation}
{\cal Z}_N =\int dq_1\dots dq_N \; \prod_{i=1}^N {\cal T}(q_i,q_{i+1}) \ , \label{Z2}
\end{equation}
recalling that $q_{N+1}\equiv q_1$.
With this notation, the (configurational) free energy density
\begin{equation}
f=-\frac{1}{\beta N} \log ( {\cal Z}_N )
\end{equation}
in the thermodynamic limit is promptly written as
\begin{equation}
f=-\frac{1}{\beta} \log ( \max{ \{\bar \lambda\}} ) \label{fp}
\end{equation}
where $\bar \lambda$ is the set of eigenvalues of the transfer
matrix, i.~e. the eigenvalues of the integral equation
\begin{equation}
\int dy \; {\cal T}(x,y) \phi(y) = \lambda \phi(x).
\end{equation}
The latter equation, with the substitution
\begin{equation}
\psi(x) = e^{\beta V_{p}(x)/2} \phi(x),
\end{equation}
turns out to be
\begin{equation}
\int dy \; e^{-\beta K \vert x-y\vert} \; \; e^{-\beta V_{p}(y)}
\psi(y) = \lambda \psi(x). \label{ei}
\end{equation}
The next step is performed by noticing that the operator $[
-d^2/dx^2+\beta^2 K^2 ]$ applied to $\exp(-\beta K\vert x-y \vert)$
produces a delta-function:
\begin{equation}
\left [ -\frac{d^2}{dx^2}+\beta^2 K^2 \right ] e^{-\beta K \vert
x-y\vert} = 2\beta K \delta(x-y),
\end{equation}
thus by applying the previous operator to the integral
equation~\ref{ei}, it can be transformed in a Schroedinger like
differential equation:
\begin{equation}
\left [ -\frac{d^2}{dx^2}+\beta^2 K^2 -\frac{2\beta K}{\lambda}
e^{-\beta V_{p}(x) } \right ] \psi(x) =0.
\end{equation}
This equation must be solved with the conditions that {\it i)} the
"eigenfunction" $\psi(x)$ was normalizable, and, {\it ii)} the
boundary condition (implicit in Eq.~\ref{ei}) $\psi'(0)/
\psi(0)=\beta K$ was fulfilled. In summary, the calculation of the
thermodynamic of system defined by the potential energy of the
form in Eq.~\ref{v_bur} is reduced to the solution of a
Schroedinger-like differential equation and, in particular, to the
finding the largest eigenvalue of the original integral equation
\ref{ei}. In general, as the eigenvalues are continuous and smooth
function of the parameters (among which the temperature), no phase
transitions are expected unless the two largest among them cross
each other.

\subsection{The $\zeta=0$ case.}

Let us now apply the procedure to the potential function in
Eq.~\ref{v_bur} for the case $\zeta=0$. 
We do not report the details of the calculation, as they are based on
standard techniques for solving Schroedinger equation in Quantum Mechanics
\cite{QM};
in summary the "eigenvalues" $\lambda$ are determined by the equation:
\begin{equation}
z(\lambda)=\beta K \label{zeta0}
\end{equation}
with
\begin{eqnarray}
z(\lambda)= \frac{f_1(P,Q) \sin{(Q R)}-f_2(P,Q))\cos{(Q R)}}
{f_3(P,Q))\sin{(Q R)}+ f_4(P,Q)\cos{(Q R)}}, \label{zeta02}
\end{eqnarray}
having defined
\begin{eqnarray}
f_1(P,Q)=Q^2
\\ \nonumber
f_2(P,Q)=P Q
\\ \nonumber
f_3(P,Q)=P
\\ \nonumber
f_4(P,Q)=Q \label{deff}
\end{eqnarray}
and
\begin{eqnarray}
Q(\lambda) = \sqrt{\frac{2\beta K}{\lambda} \; e^{\beta
U_o}-\beta^2 K^2}
\\ \nonumber
P(\lambda) =\sqrt{\beta^2 K^2-\frac{2\beta K}{\lambda} }.
\end{eqnarray}

The only possibility for the function $z(\lambda)$ to be real
(condition required for Eq.~\ref{zeta0} to have solution) is that
$P$ was real (if $Q$ become imaginary, $z(\lambda)$ is still
real), thus it exists a solution to Eq.~\ref{zeta0} only if $P$ is
real. Therefore, when $P$ vanishes, the eigenvalues $\lambda$
disappear (more specifically, disappear the eigenvalues of the
discrete spectrum, and only those of the continuum spectrum
remain), and the (configurational) free energy is discontinuous.
For each temperature, the condition $P(\lambda)=0$ is fulfilled
for a "critical" $\lambda$, given by:
\begin{equation}
\lambda_c=\frac{2}{\beta K}.
\end{equation}
Thus, the equation for the largest eigenvalue at the "critical"
point is given by $z(\lambda_c)=\beta_c K$, or
\begin{equation}
\sqrt{e^{\beta_c U_o} -1 } \tan{\left [ \beta_c K R
\sqrt{e^{\beta_c U_o} -1 } \; \right ]}= 1,
\end{equation}
which gives us the required equation for the critical (inverse)
temperature $\beta_c$.  This equation can be rearranged,
introducing the control parameter $\xi=KR/U_0$, as:
\begin{equation}
\xi = \frac{1}{\beta_c U_o }\frac{1}{\sqrt{e^{\beta_c U_o} -1 }}
\arctan{\left [ \frac{1}{\sqrt{e^{\beta_c U_o} -1 }} \; \right
]}.\label{equ}
\end{equation}
The plot of the critical temperature (in reduced units $k_BT/U_o$)
as a function of $\xi$ is reported 
as full line in Fig.~3.

\begin{figure}[t]
\includegraphics[width=.5\textwidth]{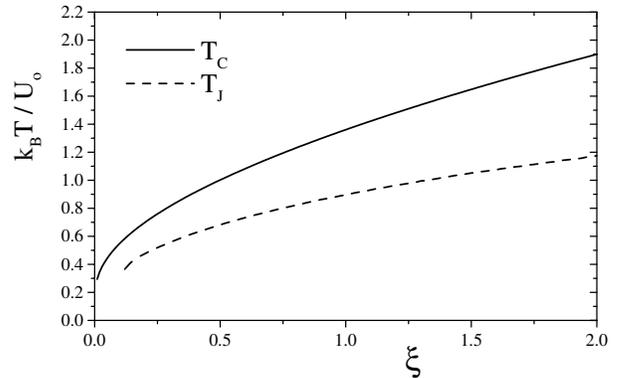}
\caption{Full line: critical temperature (in reduced units $k_BT_c/U_o$) as a
function of the control parameter $\xi$ from Eq.~\ref{equ} for the 
$\zeta$=$0$ case. 
Dashed line: temperature $T_J$ (in reduced units) at which the ``underlying saddle''
jumps from minimum to saddle as a function of $\xi$ (see Sec. V B).}
\label{fig_3}
\end{figure}

The phase transition is of the localization-delocalization type.
The particles, kept together by the $K\vert x-y \vert$ term of the
potential, for $T<T_c$ are pinned close to the square well, while,
for $T>T_c$ are delocalized in the $q$-axis.

A simple calculation leads to the value of the critical energy
$v_c$ (the equilibrium energy $v(T)$ at the transition point
$v_c=v(T_c)$). From Eq.~\ref{fp}, we have

\begin{equation}
v(T)=\frac{\partial (\beta f)}{\partial \beta} =
-\frac{\lambda'(\beta)}{\lambda(\beta)} \label{ff}
\end{equation}
where $\lambda(\beta)$ is the solution of Eq.~\ref{zeta0}. Close
to the critical point, $\lambda(\beta)=2/\beta K$, thus
$\lambda'(\beta)/\lambda(\beta)=1/\beta$ and
\begin{equation}
v_c= k_B T_c
\end{equation}
independently from the value of $\xi$.

\subsection{The $\zeta \ne 0$ case.}

\begin{figure}[t]
\includegraphics[width=.5\textwidth]{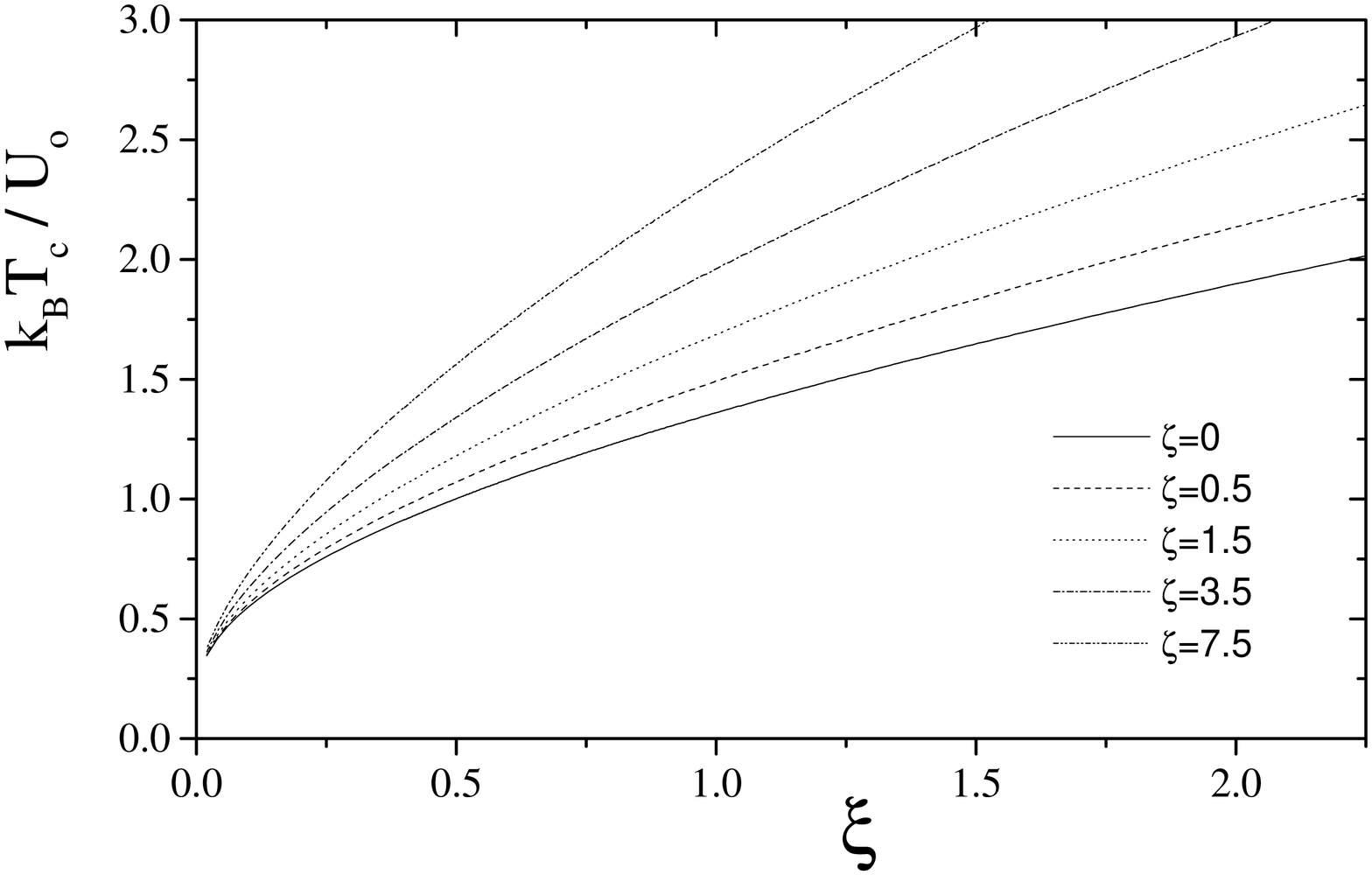}
\caption{Critical temperature (in reduced units $k_BT_c/U_o$) as a
function of the control parameter $\xi$ from Eq.~\ref{equu} for
the indicated $\zeta$ value: $\zeta=0$ (full line), $0.5$ (dashed
line), $1.5$ (dotted line), $3.5$ (dash-dotted line) and $7.5$
(dash-dot-dotted line). } \label{fig_4}
\end{figure}

The calculation for the case of generic $\zeta$ values is
quite similar to the previous one. 
Also in this case, the eigenvalues
$\lambda$ are determined by an equation like Eq.~\ref{zeta0},
$z(\lambda)=\beta K$, 
with $z(\lambda)$ again given by Eq.~\ref{zeta02}
and with the $f_n(P,Q)$ functions ($n$=$1,\dots,4$) given by
\begin{eqnarray}
f_1(P,Q)&=&P \left [ Q^2 \cosh(P R \zeta) - P^2 \sinh(P R \zeta)
\right] \nonumber
\\ \nonumber
f_2(P,Q)&=&P^2 Q \exp{(P R \zeta)}
\\ \nonumber
f_3(P,Q)&=& \left [ P^2 \cosh(P R \zeta) - Q^2 \sinh(P R \zeta)
\right]
\\
f_4(P,Q)&=&P Q \exp{(P R \zeta)}. \label{deff2}
\end{eqnarray}
Obviously, Eqs.~\ref{deff2} recover Eqs.~\ref{deff} in the $\zeta
\rightarrow 0$ limit. The same considerations on the reality of
$P(\lambda)$ reported above apply to Eq.~\ref{deff2}. 
Thus the condition $P(\lambda)=0$ define the critical
value of the eigenvalue, $\lambda_c=2/\beta K$,and the equation
for the critical temperature ($z(\lambda_c)=\beta_c K$) becomes:
\begin{eqnarray}
&& \sqrt{e^{\beta_c U_o} -1 } \sin{\left ( \beta_c K R
\sqrt{e^{\beta_c U_o} -1 } \; \right )} \times \\ && \times \big [
\cos{\left ( \beta_c K R \sqrt{e^{\beta_c U_o} -1 } \; \right )}-
\nonumber
\\ && - \beta_c K R \zeta \sqrt{e^{\beta_c U_o} -1 } \sin{\left (
\beta_c K R \sqrt{e^{\beta_c U_o} -1 } \; \right )} \big ]^{-1}=
1. \nonumber
\end{eqnarray}
Similar to the $\zeta=0$ case, this equation can be rearranged,
introducing the control parameter $\xi$, as:
\begin{equation}
\xi = \frac{1}{\beta_c U_o }\frac{1}{\sqrt{e^{\beta_c U_o} -1 }}
\arctan{\left ( \frac{1}{\sqrt{e^{\beta_c U_o} -1 }} \;
\frac{1}{1+\beta_c U_o \xi \zeta } \;\right )}.\label{equu}
\end{equation}
At variance with Eq.~\ref{equ}, this equation cannot be cast in
the form $\xi=\xi(\beta_c)$, thus it must be solved numerically to
plot the critical temperature as a function of the control
parameter $\xi$. This plot is reported in Fig.~\ref{fig_4} for
different values of $\zeta$.

\begin{figure}[t]
\includegraphics[width=.5\textwidth]{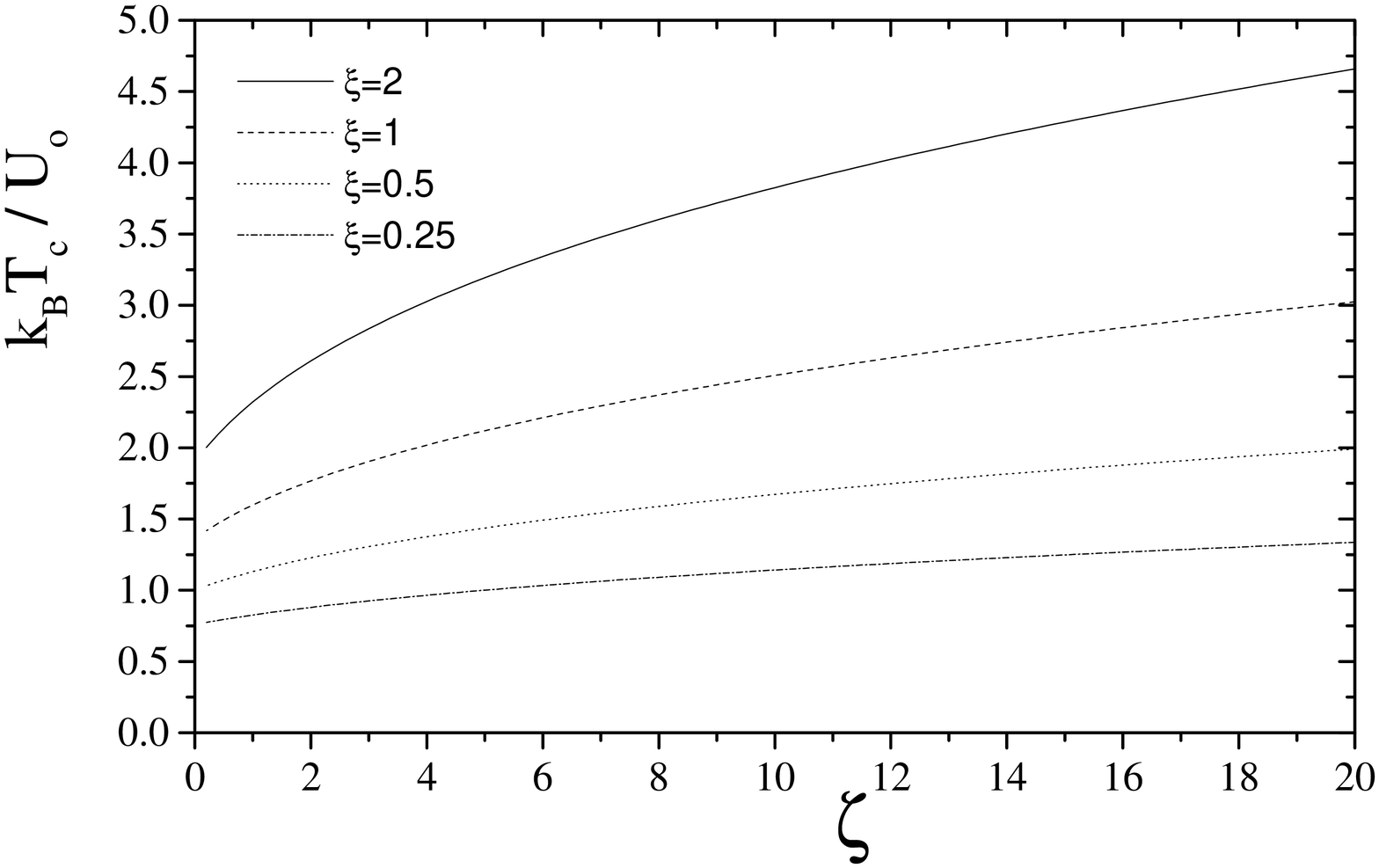}
\caption{Critical temperature (in reduced units $k_BT_c/U_o$) as a
function of the control parameter $\zeta$ from Eq.~\ref{equu} for
the indicated $\xi$ value: $\xi=2$ (full line),$1$(dashed line),
$0.5$ (dotted line), $0.25$ (dash-dotted line).  } \label{fig_5}
\end{figure}

As can be observed in Fig.~\ref{fig_4}, on increasing $\zeta$,
i.~e. on displacing the position of the square well towards high
value of the coordinate, the critical temperature, for a given
$\xi$ value, increases, expanding the amplitude of the "cold"
(localized or pinned) phase. This can be better seen in
Fig.~\ref{fig_5}, where the $\zeta$ dependence of the critical
temperature is reported for some values of $\xi$. We conclude this
section noticing that the phase transition actually exists for all
the value of $\zeta$, and in the limit of $\zeta \rightarrow
\infty$, the critical temperature goes {\it without
discontinuities} to infinity. 
Therefore, we are lead to conclude that the model investigated in Ref.~\cite{kastner} to
demonstrate the unattainability of a purely topological criterion for the
existence of a phase transition is a ``borderline'' model,
in which the phase transition can be thought to be present at 
``$T$ infinity'' (even though the precise meaning of this statement
is not well defined). Then, to the same
topology (as we will see in the next section, the topology of the
potential function in Eq.~\ref{gburk} does not depend on the value of
$\zeta$) always corresponds a phase transition.

\begin{figure}[t]
\includegraphics[width=.5\textwidth]{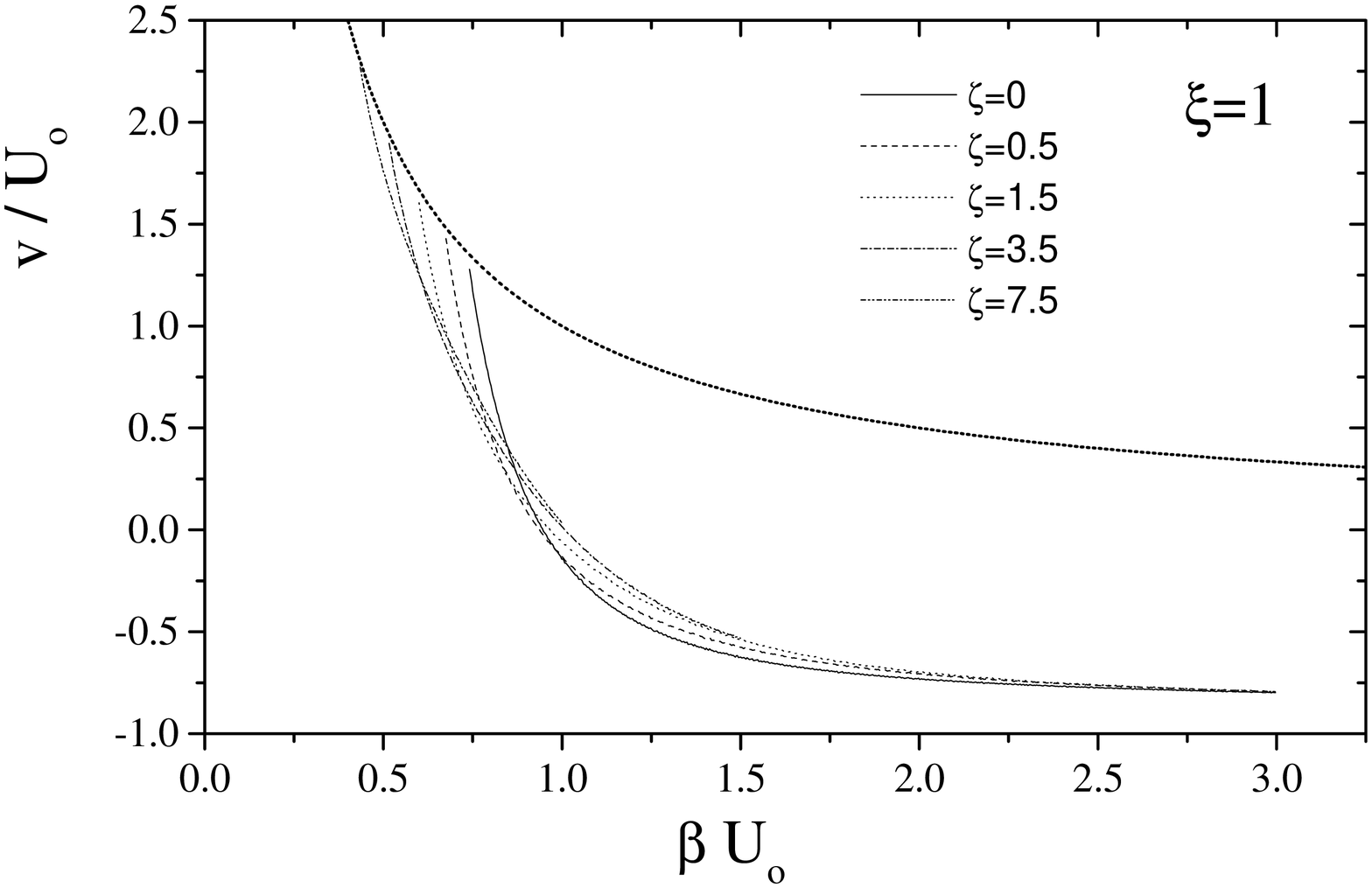}
\caption{Inverse temperature dependence (in reduced units $\beta
U_o$) of the equilibrium potential energy for $\xi=1$ and for the
indicated $\zeta$ value: $\zeta=0$ (full line), 0.5 (dashed line),
1.5 (dotted line), 3.5 (dash-dotted line) and 7.5 (dash-dot-dotted
line). The thick-dotted line represents the $\beta$ dependence of
the potential energy in the high temperature phase.} \label{fig_6}
\end{figure}

To discuss the question of the {\it coincidence} (or not) of the
critical energy with the topological discontinuity, we need to
calculate $v_c(\xi,\zeta)$. Following the same argument reported
for the case $\zeta=0$ we conclude that the critical potential
energy depends on $\xi$ and $\zeta$ only through $T_c$: $v_c=k_B T_c$. 
As an example, in Fig.~\ref{fig_6} we report
the caloric curve $v(T)$ as a function of the inverse
temperature for different $\zeta$ value and for $\xi=1$. For all
the $\zeta$ values, on the low-$\beta$ side the curves end at the
points ($\beta_c$, $v_c$); these points are aligned along the
$v_c(\beta_c)$ line (thick dotted line) given by
$v_c(\beta_c)=1/\beta_c$.

\subsection{Topology}
The analysis of the topological properties of the 
Burkhardt model is reported by Kastner in Ref. \cite{kastner}.
He analyzed only the two limiting cases of confining
and non-confining models, corresponding in our notation
to $\zeta$=$0$ and $\zeta$=$\infty$ respectively.
He found that a topology change is present in both cases,
even if not really equal in ``strength''.
The value of the potential energy at which the topological change
appears is $v_{\theta}$=$0$, irrespective of the considered model.
One can easy generalize the above analysis to the general case
with arbitrary $\zeta$, and conclude that the topological change is 
always located at energy $v_{\theta}$=$0$. 
It is worth noting that the energy at which topological change appears
is lower than the thermodynamic transition energy: $v_c > v_{\theta}$
(see Fig. \ref{fig_6}).
We will further discuss this issue in Sec. V, after having described 
the thermodynamics and topology of the PB and SPB models.
 
\section{Peyrard-Bishop model}

\subsection{Thermodynamics}
\begin{figure}[t]
\includegraphics[width=.5\textwidth]{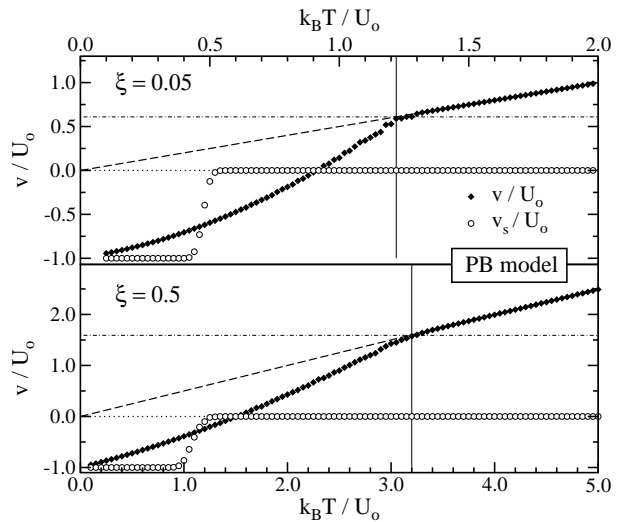}
\caption{
Temperature dependence (in reduced units $k_B T /U_o$) of the
equilibrium potential energy $v$ (full symbols)
and energy $v_s$ of underlying saddles (open symbols)
for the PB model defined by Eq. \ref{model1} with
$\xi$=$0.05$ (upper panel) and $\xi$=$0.5$ (lower panel).
Also indicated in the figure are the values of
topological change energy $v_{\theta}$ (horizontal dotted lines),
phase transition energy $v_c$ (horizontal dot-dashed lines)
and transition temperature $T_c$ (vertical full line) 
for $\xi$=$0.05$ ($v_{\theta}/U_o$=$0$, $v_c/U_o\simeq0.61$ and 
$k_BT_c/U_o \simeq 1.22$)
and for $\xi$=$0.5$ ($v_{\theta}/U_o$=$0$, $v_c/U_o \simeq1.59$ and 
$k_BT_c/U_o \simeq 3.20$).
Dashed lines are the $T$-dependence of the potential energy in the
high $T$ phase: $v(T)$=$k_B T/2$.
}
\label{fig_7}
\end{figure}

The thermodynamics of the Peyrad-Bishop model 
(defined in Eq.s \ref{v_pb}, \ref{model1}) can be studied using 
transfer matrix techniques, as the Burkhardt model 
described in the previous section.
However, in this case approximated methods have to be considered
in order to obtain a corresponding Schroedinger 
like differential equation.
In the region $\xi \gg 1$ and
temperature window $U_o \ll k_B T \ll \xi U_o$
the classical statistical mechanics problem is mapped
to the quantum Morse oscillator problem \cite{pey2,theo}.
Similarly to the case of Burkhardt potential,
the presence of a second order phase transition for the Peyrad-Bishop model
is signaled by the bounded-unbounded transition of the
lower state in the corresponding quantum problem.
In the above range of $\xi$ and $T$, Peyrard and Bishop obtained an analytical
expression for the transition temperature
$k_B T_c / U_o$=$4\sqrt{\xi}$
and transition energy 
$v_c / U_o$=$k_B T_c/2U_o$=$2\sqrt{\xi}$.
For generic $(\xi,T)$ values, only numerical results can be used to
infer the existence and location of a phase transition. In Fig.
\ref{fig_7} we report the temperature dependence of the potential
energy per particle $v=V/N$ (full symbols) of the PB model for
two different values of $\xi$: $0.05$ (upper panel) and $0.5$ (lower
panel). Also reported in the figure are the energy $v_c$ (dot-dashed
line) and temperature $T_c$ (full line) of the phase transition point: 
$v_c/U_o \simeq 0.61$ and $k_B T_c/U_o \simeq 1.22$ for $\xi$=$0.05$,
$v_c/U_o \simeq1.59$ and $k_B T_c/U_o \simeq3.20$ for $\xi$=$0.5$. 
Dashed lines are
the $T$-dependence of the potential energy in the high $T$ phase:
$v(T)$=$k_B T/2$.

\subsection{Topology}
The topology of the Peyrard-Bishop model is studied in the paper
of Grinza and Mossa \cite{gri_mos}.
A topological change is found at the energy value 
$v_{\theta}$=$0$, corresponding to a
topological change in the hypersurfaces $\Sigma_v$ varying $v$:
from a close hypersurface for $v < v_{\theta}$ to an open one for
$v \geq v_{\theta}$ \cite{gri_mos}.
In Fig. \ref{fig_7} the value of $v_{\theta}$ is 
indicated by an horizontal dotted line.
We note that, also in this case, the topological discontinuity is lower in energy
than the thermodynamic one: $v_c > v_{\theta}$.


\section{Symmetric Peyrard-Bishop model}

\subsection{Thermodynamics}
\begin{figure}[t]
\includegraphics[width=.5\textwidth]{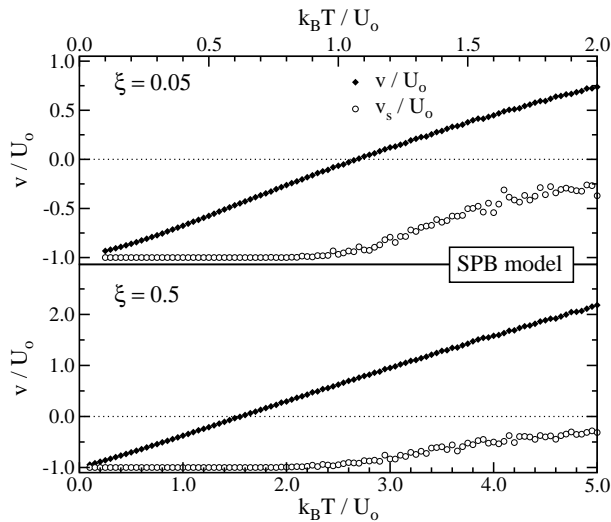}
\caption{
Temperature dependence (in reduced units $k_B T /U_o$) of the
equilibrium potential energy $v$ (full symbols)
and energy $v_s$ of underlying saddles (open symbols)
for the SPB model defined by Eq. \ref{model2} with
$\xi$=$0.05$ (upper panel) and $\xi$=$0.5$ (lower panel).
Also indicated in the figure is the value of
the topological change energy $v_{\theta}/U_o$=$0$ (dotted line)
for both cases.
}
\label{fig_8}
\end{figure}

The Symmetric Peyrard-Bishop model defined by Eq.s \ref{v_pb},
\ref{model2} does not exhibit phase transition at finite $T$. 
This can be view from the fact that
there is always a bound state in the corresponding quantum problem,
in analogy with the non-confined Burkhardt model
\cite{bur,cuesta}.
In Fig. \ref{fig_8} the
same quantities as in the PB case are reported for the SPB model:
energy $v$ (full symbols) for $\xi$=$0.05$ (upper panel) and $\xi$=$0.5$
(lower panel). It is evident in this case the absence of a phase
transition at finite $T$ (in the $T$-range investigated).

\subsection{Topology}
Following a similar argument as in Ref. \cite{kastner}, one can
see that also in the SPB case one has a topological change  at exactly the
same energy level as in the PB model $v_{\theta}$=$0$
(even if not identical in strength to the previous one). 
We refer to the papers in
Ref.s \cite{kastner,gri_mos} for a more detailed discussion of the
topology.
In Fig. \ref{fig_8} the value of $v_{\theta}$ is indicated by an 
horizontal dotted line.

\section{Underlying saddles}

In this section we study the properties of the stationary points visited 
by the systems. The concept of ``{\em underlying saddles}'' was first
introduced in the study of glassy disordered systems \cite{sad_lj,sad_cav,cavagna}
to better understand the topological counterpart of the dynamic transition 
taking place in these systems.
Recently, it has been applied also in the analysis of models that exhibit
thermodynamic phase transitions, in order to emphasize the role of topological 
changes at the 
``{\em underlying saddles}'' energy in driving the phase transition 
\cite{ktrig,ktrig2,phi4}.

Here we apply the same methodology to investigate
the one dimensional systems introduced before.
Let start with the models having a continuous potential energy function,
the PB and SPB models, which allow for the usual definition of stationary points.
At the end of the section we will extend the argument to the discontinuous case 
of Burkhardt model.

\subsection{Peyrard-Bishop and Symmetric Peyrard-Bishop models}

There are only two stationary points in the potential energy
hypersurface of both models: a minimum located at 
$q_1$=$q_2$=$\dots$=$q_N$=$0$
and a saddle (with degenerate Hessian matrix) at
$q_1$=$q_2$=$\dots$=$q_N$=$\infty$ \cite{gri_mos}. 
In order to associate one of the two stationary points
to each instantaneous configuration of the system, 
we used a similar trick as in the analysis of 
glassy systems \cite{sad_lj} or mean-field models
\cite{ktrig,ktrig2,phi4}.
In the latter one minimized the pseudo-potential $W$=$|\nabla V|^2$
during the dynamic evolution at different temperatures,
so introducing a map from equilibrium energy levels to saddles energy levels:
${\cal{M}}\!: v\to {\cal{M}}(v)\equiv v_s$.
Due to the peculiarity of the present models,
where the saddle point is ``infinitely'' far from each equilibrium configuration,
we decided to apply the $W$ minimization method
in a two steps procedure:
{\em i}) first we minimized the $W_{int}$ quantity defined using the interaction
potential part of $V$,  $W_{int}$=$|\nabla V_{int}|^2$,
where $V_{int}$=$\sum_{i=1}^{N} \frac{K}{2} (q_{i+1}-q_i)^2$;
{\em ii}) then we minimized the $W_p$ defined using the on site
potential $W_{p}$=$|\nabla V_{p}^{(2,3)}|^2$.
This procedure ensures that the point reached is a true stationary point,
i.e. the minimum or the saddle.
Obviously, this is a quite arbitrary definition of basins of attraction
of stationary points. As said in the introduction, the robustness of the
results with respect to the possible choices of definition of a saddle
basin of attraction is still an open problem.

\begin{figure}[t]
\includegraphics[width=.5\textwidth]{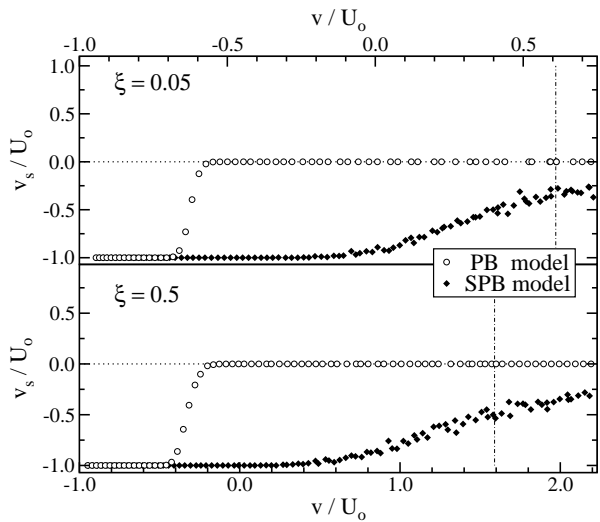}
\caption{
Map ${\cal M}\!: v \to v_s$ defined minimizing the pseudopotential $W$=$|\nabla V|^2$
in the PB and SPB models for $\xi$=$0.05$ (upper panel) and
$\xi$=$0.5$ (lower panel).
Also reported are the corresponding $v_{\theta}$ (dotted lines) and $v_c$ (dot-dashed lines)
for $\xi$=$0.05$ ($v_{\theta}$=$0$, $v_c\simeq0.61$) and $\xi$=$0.5$ ($v_{\theta}$=$0$, $v_c\simeq1.59$),
evidencing the identity ${\cal M}(v_c)$=$v_{\theta}$.
}
\label{fig_9}
\end{figure}

In Fig. \ref{fig_7} the temperature dependence of the energy $v_s$
(open symbols) of underlying saddles 
is shown for the case $\xi$=$0.05$ (upper panel)
and $\xi$=$0.5$ (lower panel) in the PB model.
The remarkable fact is that at $T_c$ (vertical full line in Fig. \ref{fig_7}) 
the identity $v_s$=$v_{\theta}$ holds. 
The map ${\cal M}(v)$ is shown for PB model (open symbols)
in Fig. \ref{fig_9}
for the two cases $\xi$=$0.05$ (upper panel) and $\xi$=$0.5$ (lower
panel). One observe that, as before pointed out, one has
${\cal M}(v_c)$=$v_{\theta}$ for both $\xi$ values. 
The fact that $v_s(T)$ in Fig.~\ref{fig_7}, as well as ${\cal M}(v)$
in Fig.~\ref{fig_9}, has a ``smooth'' transition between its low $T$ (or $v$) 
and high $T$ (high $v$) regions is most likely due to a finite size effect
($N$=$500$ here) and both $v_s(T)$ and ${\cal M}(v)$ will probably tend towards 
a step function in the thermodynamic limit.
The previous finding
indicates that the relevant quantity to consider when we are
looking for topological changes related to a phase transition is
the underlying stationary point energy, obtained trough a map from
the critical level $v_c$.
It is worth noting that the map ${\cal M}$ is constant
(${\cal M}(v)$=$v_{\theta}$) for a broad range of values,
also below $v_c$,
at variance with other cases where around the transition point
the properties of visited saddles change \cite{XY,ktrig,ktrig2,phi4}.
One can conjecture that the flatness of ${\cal M}(v)$
is a pathology of these one-dimensional models, 
that have a number of stationary points that is not extensive in $N$
(actually there are only $2$ stationary points).
 
In Fig \ref{fig_8} we report the same quantities $v_s$ as before (open symbols),
now for the SPB model, with $\xi$=$0.05$ (upper panel) and $\xi$=$0.5$
(lower panel). In this case no phase transition is present, and indeed the
topological singularity is never visited, $v_s(T)<v_\theta$
for each finite temperature ($T<\infty$).

\subsection{Burkhardt  model}

To apply the analysis of the previous section also to the Burkhardt model, one has
to find a suitable definition of ``saddles'' and of ``basin of attraction of a saddle''
for a discontinuous potential.
One possibility is the following: we first minimize the interaction potential 
$V_{int} = \sum_{i=1}^N K |q_{i+1}-q_i|$, which is equivalent 
to put all the $q_i$ equal
to the center of mass coordinate $\bar q=N^{-1}\sum_i q_i$. 
If $\bar q$ lies in the well of the potential, {\it i.e.} $\bar q \in [L,L+R]$, we
will associate the ``minimum'' to the initial configuration, otherwise we will
associate it to the ``saddle'' (we use this terminology by analogy with the PB model).
It is clear that the average energy of the ``underlying saddles'' is simply the
average of the on-site energy of the center of mass coordinate, 
\begin{equation}
v_s(T)=\langle V^{(1)}_p(\bar q) \rangle_T \ .
\label{cm}
\end{equation}
In the thermodynamic limit the center of mass $\bar q$ is peaked around its mean value
and then we can substitute the right hand side of Eq. \ref{cm} with 
$V^{(1)}_p(\langle \bar q \rangle)$, a quantity that can be explicitly computed.
To determine $\langle \bar q \rangle$ we can use the distribution probability 
$|\phi(x)|^2$, where $\phi(x)$=$e^{-\beta V_{p}(x)/2} \psi(x)$ and $\psi(x)$ is the eigenfunction
of the transfer matrix operator corresponding to the maximum eigenvalue \cite{bur} (see Sec. II A).
We note that the saddle energy $v_s(T)$ is a step function, equals to the minimum energy $-U_o$ 
when $\langle \bar q \rangle$ lies inside the square well and equals to the saddle 
energy $0$ otherwise. 
The temperature $T_J$ at which the visited ``underlying saddle'' jumps from minimum to saddle
is shown in Fig. \ref{fig_3} (dashed line) as a function of the 
parameter $\xi$ for the $\zeta$=$0$ case.
It is worth noting that the temperature $T_J$ lies always below the thermodynamic transition
temperature $T_c$ (in analogy with the PB model, see Fig. \ref{fig_7}).
The same happens for all values of $\xi$.
Therefore, also for the Burkhardt case, at the transition temperature $T_c$
the ``underlying saddles'' lie at an energy equal to the topological discontinuity energy
$v_{\theta}$, i.e. ${\cal M}(v_c)$=$v_{\theta}$.

\section{Conclusions}                 

Studying two particular one dimensional models discussed in the
recent literature \cite{kastner,gri_mos}
(Burkhardt model in the confining and non-confining version,
Peyrard-Bishop model and its non-confining counterpart),
we have focused on the relationship between phase transitions and
topological changes, 
recently proposed in the literature \cite{cccp,fps,cpc,fra_pet}.
In these models, a topological
singularity at a given energy value $v_\theta$(=0) is always found;
however, {\it i)} in the confining version
a phase transition is found but the critical energy is $v_c > v_\theta$ 
\cite{gri_mos}; {\it ii)} in their non-confining version there is no phase transition 
at any finite temperature \cite{kastner}.

These results generated confusion as {\it i)} was
interpreted as a confirmation of the {\em strong topological hypothesis} of
Pettini {\it et al.} \cite{nota} 
while {\it ii)} was considered as an evidence for the 
unattainability of a purely topological criterion
for detecting phase transitions, 
although demonstrated only for the particular non-confining
one dimensional models.

Exploiting the concept of ``{\em underlying
stationary points}'' defined through a generalization 
of the methods used in the glassy literature 
(minimization of the pseudopotential $W$=$|\nabla V|^2$), 
we have defined a map ${\cal M}:v\to
v_s$ from energy level $v$ of $V$ to stationary points, with energy
$v_s$. 
We have shown that: {\it i)} in the confining case, where the phase
transition is present, one has ${\cal M}(v_c)=v_\theta$, in agreement
with the {\em weak topological hypothesis};
{\it ii)} in the non-confining case, where the phase transition is not
present at finite temperature 
(as the transition temperature goes continuously to infinity when
the confining wall is removed) the energy of the underlying saddles is
always {\it below} the topological singularity, i.e. $v_s(T)<v_\theta \, , \, \forall T$;
the singular point $v_\theta$ is indeed visited for $T\rightarrow \infty$,
consistently with the observation that the critical temperature is ``infinite''
in the non-confining case. 

The {\it weak topological hypothesis} appears as a possible
framework to fit the results that recently appeared in the
literature on all the different models investigated so far.
Within this hypothesis three different scenarios are possible:
\begin{enumerate}
\item If there is no topological singularity $v_\theta$, a phase
transition is not possible; this is consistent with the hypothesis
of Pettini {\it et al.}: topological singularities are {\it necessary}
conditions for a phase transition to take place.
\item If there is a topological singularity at energy $v_\theta$, 
a phase transition is also
present {\it if and only if} there exist a temperature $T_c$ such that 
$v_s(T_c) = v_\theta$
(or equivalently an energy $v_c$ such that ${\cal M}(v_c)=v_\theta$).
\end{enumerate}
The above findings seem to indicate that, at least for the particular models investigated,
a sufficiency criterion for the phase transition to take place requires 
the introduction of a statistical measure:
thus, we believe that the statement of Kastner~\cite{kastner}
concerning the unattainability of a purely topological criterion
for detecting phase transitions is indeed correct,
even though in Ref.~\cite{kastner} it has been derived using 
a ``borderline'' model (see Section II C).


Let us conclude with two remarks:
{\it i)} as already stated, the definition of the map ${\cal M}$ is
not unique, different definitions giving (slightly) different results.
Thus, the {\it weak topological hypothesis} contains in its formulation
an ambiguity and must be regarded only as a {\it practical} tool, at
least at this stage of comprehension;
{\it ii)} nevertheless, we hope that this approach can be of interest
for the numerical investigation of systems of ``mesoscopic'' size
(e.g. proteins and large molecules),
i.e. such that the number of degrees of freedom is not large enough 
to allow to detect the presence of a phase transition using
standard techniques.

We thank M.~Pettini and M.~Kastner for helpful comments and suggestions.

\end{document}